\documentclass[preprint,11pt]{elsarticle}

\usepackage{amssymb}

\biboptions{sort&compress}

\newcounter{bla}

\usepackage{amsfonts}
\usepackage{amsmath}
\usepackage{amssymb}
\usepackage{amsthm}
\usepackage{graphicx}
\usepackage{float}
\usepackage{bigints}
\usepackage[config,labelfont=sc,textfont=sl]{caption,subfig}
\usepackage[british]{babel}
\usepackage{bm}
\usepackage{xcolor}
\usepackage{ulem}
\newcommand{\bv}[1]{\boldsymbol{#1}}

\DeclareMathOperator*{\real}{Re}
\DeclareMathOperator*{\imag}{Im}

\newcommand{\nc}{\newcommand}
\nc{\nn}{\nonumber}
\nc{\XYZ}{}

\journal{Computer Physics Communications}
\makeatletter
\def\ps@pprintTitle{%
  \let\@oddhead\@empty
  \let\@evenhead\@empty
  \let\@oddfoot\@empty
  \let\@evenfoot\@oddfoot
}
\makeatother

\begin{document}

\begin{frontmatter}



\title{An improved argument principle root-search method for modes of slab waveguides, optical fibers, and spheres}


\author[a]{Sravya Rao\corref{author}\fnref{1}}
\author[a]{Parry Y.\ Chen\fnref{1}}
\author[a]{T. Grossinger}
\author[a]{Yonatan Sivan}

\cortext[author] {Corresponding author -
\textit{E-mail address:} sravya@post.bgu.ac.il}
\fntext[1]{Equal contributors to this update. }
\address[a]{School of Electrical and Computer Engineering, Ben-Gurion University, Israel}

\begin{abstract}
We update our root-search method for transcendental equations. Our method is globally convergent and is guaranteed to locate all complex roots within a specified search domain, since it is based on Cauchy’s residue theorem. We extend the implementation to treat the dispersion relations of slab waveguides and the resonances of a sphere, in addition to step-index fibers. We also implement other improvements, such as to the contour selection procedure {\XYZ and using non-dimensional search variables,} to ensure the method remains reliable even in challenging parameter regimes. {\XYZ We also extend the algorithm to identify leaky modes in terms of propagation constant eigenvalue modes, revealing, to the first time to our knowledge, a discontinuity across the light line in the dispersion plot.}
\end{abstract}

\begin{keyword}
Complex root search; Argument principle method; Optical waveguide dispersion relation.
\end{keyword}
\end{frontmatter}
{\bf NEW VERSION PROGRAM SUMMARY}\\
\begin{small}
\noindent
{\em Program Title:} disproots (Dispersion roots)  \\
{\em CPC Library link to program files:} (to be added by Technical Editor) \\
{\em Developer's repository link:} (if available) \\
{\em Code Ocean capsule:} (to be added by Technical Editor)\\
{\em Licensing provisions(please choose one):} CC BY NC 3.0  \\
{\em Programming language:} MATLAB                                  \\
{\em Journal reference of previous version:} Comput. Phys. Comm. 214 (2017) 105.                  \\
{\em Does the new version supersede the previous version?:}  Yes  \\
{\em Reasons for the new version:} Even though the algorithm is theoretically guaranteed to locate all roots, our original implementation~\cite{Parry-wire-principal_value} is still liable to miss some of the roots at certain parameter regimes in the context of the optical waveguide dispersion relation. In this update, we make the algorithm more reliable. The other reason for this update is to extend the implementation of our algorithm to treat the dispersion relations of other waveguide geometries and applicability in determining the {\XYZ leaky} modes in the context of propagation constant modes. \\

{\em Summary of revisions:} To ensure the method is reliable even in challenging parameter regimes, we make several revisions in the contour selection strategy like enlargement of the contour for lower-order roots, {\XYZ adapting the simple circular contours used previously~\cite{Parry-wire-principal_value} by} elliptically-shaped contours and employing adaptive scheme for higher-order roots. Additionally, we also rescale the search variables in the context of the propagation constant when the spacing between roots is large. We further demonstrate the extension of the method to slab \& sphere geometries, and computation of {\XYZ leaky} modes.\\

{\em Nature of problem:} Locating the complex roots of a general transcendental equation is often a non-trivial task. Iterative methods such as Newton’s method face numerous difficulties because the existence of roots with narrow and otherwise difficult attraction basins requires very accurate initial guesses to locate. Robust location of a complete set of roots thus becomes problematic. In optics, modes of a circular fiber are obtained from a transcendental equation, the dispersion relation. Several recent advancements in optics have necessitated its robust solution, including the fabrication of high index contrast fibers and analytical methods that expand radiating sources using eigenmodes.\\

{\em Solution method:} We employ the argument principle method, a robust globally convergent method guaranteed to locate all roots in the specified search domain. It is based on the Cauchy residue theorem, and projects the locations of the roots on to a polynomial basis. Unlike previous implementations of the argument principle method [2,3] and related methods [4], our implementation has two features vital for solving the fiber dispersion relation. It allows isolated singularities within the search domain, and allows the search domain to approach arbitrarily close to branch points without experiencing failure. Furthermore, our simple MATLAB implementation is designed to be easily modified and integrated for a variety of applications.\\

{\em Additional comments including restrictions and unusual features:} The specified search domain must be meromorphic, in other words be complex analytic containing at most isolated singularities. The locations and orders of these singularities must be known analytically, so we describe how they are determined for our fiber dispersion relation. Branch points and branch cuts must be avoided, though the search domain may be arbitrarily close. We exploit knowledge of our dispersion relation, specifically that roots are located close to singularities, to simplify the method.\\
   \\

\end{small}

\section{Introduction}
\label{}
In this update, we revisit our original algorithm~\cite{Parry-wire-principal_value}, where we implemented a robust, globally convergent argument principle root-search subroutine for transcendental equations. The algorithm is guaranteed to locate all roots within a specified search domain of the complex plane, since it is a globally convergent algorithm based on Cauchy’s residue theorem. We applied the algorithm to treat the step-index optical fiber dispersion relation, i.e., to find dispersion curves ($\omega$-frequency vs $\beta$-propagation constant~\cite{Yeh-book,okamoto-book}, or $\beta$-modes in short) given a user-specified fiber radius and material dispersion relation. Free-to-use code was published alongside the original article~\cite{Parry-wire-principal_value}. The algorithm found the roots up to any desired order with minimal human supervision, and without any concerns of divergence with increased number of iterations~\cite{bai2013efficient}.  The implementation can also search for permittivity as the eigenvalue, which is useful for defining the normal modes of open, unbounded systems~\cite{Bergman-spectral-decomposition-book,Bergman_Stroud_PRB_1980,Sandu,Nitzan_hybridization,Gersten_Nitzan_hybridization,Russians-permittivity-modes-book,Li:2004cq,Chong_Stone_SALT,Schnitzer_PRB_2015,Schnitzer_Proc_R_Soc_A_2016,Farhi_Bergman_PRA_2016,GENOME,Parry-Egor-re-expansion,Kossowski-Chen,Parry-GALM,Gilles_hybridization,Gilles_graphene_nanoislands,Forestiere2016,Forestiere2017,Forestiere2018,Pascale2019,Sravya_josab_2022,frequency_expansion_GENOME_josab_2022,Boag_2022}.

\begin{figure}[t]
   \begin{center}
   \includegraphics[scale=1.5]{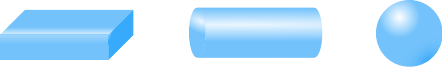}
   \end{center}
   \caption{The three geometries treated in this update: a slab waveguide with infinite translational symmetry in two dimensions, a circular fiber infinitely extended in one direction, and a sphere.}
   \label{fig:geometries}
\end{figure}

The first purpose of this update is to enhance the reliability of our implementation in the context of optical waveguide dispersion relations.
{\XYZ Even though the algorithm is theoretically guaranteed to locate all roots, our original implementation~\cite{Parry-wire-principal_value} is still liable to miss some of the roots.}
There are two possible reasons for this. Firstly, if the search domain is not specified to be large enough to encompass all the roots, then roots that stray deep into the complex plane will be missed. This can occur under certain parameter regimes. Secondly, the algorithm might fail due to the limitations of finite precision arithmetic when the search variable becomes very small or very large. In this update, we revise the algorithm to resolve these issues, yielding a more reliable implementation. {\XYZ These improvements are described in Section~\ref{sub:revision}. }

The {\XYZ second purpose} of this update is to demonstrate its applicability to compute the {\XYZ {\em leaky}} optical waveguide ($\beta$) modes, i.e., those for which the search variable is a complex propagation constant $\beta$; previously, we studied such modes only when the search variable was the (eigen-) permittivity. Such modes are encountered also in the context of the Schr\"odinger equation in the description of electron states whose energies are above the vacuum level~\cite{Moiseyev-book}; the optical formulation adopted below reduces to this case for a fixed value of the frequency. {\XYZ Details of the leaky mode computation are presented in Section~\ref{sub:Leaky}}.

The {\XYZ third purpose} of this update is to extend the implementation of our algorithm to compute the dispersion relations of slab waveguide and spherical scatterers, see Fig.~\ref{fig:geometries}. 
{\XYZ These structures yield simpler transcendental equations compared to step-index optical fiber, often resulting in more regular root distributions and simpler attraction basins. 
Thus, these root searches are perhaps less challenging and may be more amenable to simpler techniques, such as the locally convergent Newton’s method. However, since the roots can lie in the complex plane, locally convergent techniques can still be unreliable, as their success depends heavily on good initial guesses. To ensure that no roots are missed, it remains beneficial to use our globally convergent algorithm. In Section~\ref{sub:other_geometries}, we demonstrate the implementation of the algorithm for a symmetric slab waveguide and for spherical geometries.}

\section{Novelities}
\subsection{Revisions to algorithm}\label{sub:revision}

{\XYZ Our root-search algorithm~\cite{Parry-wire-principal_value} was originally implemented to solve the transcendental equation governing the circular fiber, defined by
\begin{equation}\label{eq:dispersion}
\begin{gathered}
f_m \equiv \left(\frac{\mu_c}{\alpha_c a}\frac{J'_m(\alpha_c a)}{J_m(\alpha_c a)} - \frac{\mu_b}{\alpha_b a}\frac{H'_m(\alpha_b a)}{H_m(\alpha_b a)}\right) \left(\frac{\epsilon_c}{\alpha_c a}\frac{J'_m(\alpha_c a)}{J_m(\alpha_c a)} - \frac{\epsilon_b}{\alpha_b a}\frac{H'_m(\alpha_b a)}{H_m(\alpha_b a)}\right) \\
- \left(\frac{m\beta}{k}\right)^2 \left(\frac{1}{(\alpha_c a)^2} - \frac{1}{(\alpha_b a)^2}\right)^2 = 0.
\end{gathered}
\end{equation}
Here, $J_m$ denotes the Bessel function of the first kind, and $H_m$ is the Hankel function of the first kind. The index $m$ represents the angular variation of the field ($\sim e^{i m \theta}$), $a$ is the radius of the circular core, $\epsilon$ and $\mu$ are the relative permittivity and permeability of the core (denoted by subscript $c$) and background ($b$), $k = \omega/c$ is the free space wavenumber, and $\beta$ is the propagation constant along the fiber axis. The in-plane propagation constants $\alpha_c$ and $\alpha_b$ differ between the core and background,
\begin{equation}
    \alpha_c^2 = k^2 \epsilon_c \mu_c - \beta^2, \quad \alpha_b^2 = k^2 \epsilon_b \mu_b - \beta^2.
    \label{eq:fiber_alphas}
\end{equation}}
{\XYZ In this implementation, the root search was carried out using either $\epsilon_c$ or $\beta$ as the search variable, with the} contour selection strategy guided by two key observations: (a) there exists one root for each singularity, counting multiplicities, and (b) these roots are often located in the vicinity of their respective singularities. We thus devised a simple strategy to enclose all the roots with a set of overlapping contours, with one contour centered on each singularity. This was specified in Eq.~(14) of our original manuscript~\cite{Parry-wire-principal_value} and demonstrated there in Fig.~2. However, the strategy can fail under certain parameter regimes. {\XYZ For instance, in the long-wavelength limit ($ka < 1$) or when the index contrast is small, 
roots tend to deviate more significantly from their associated singularities, and the contours may no longer adequately enclose them. This deviation manifests along both the real and imaginary axes of the complex plane of search variables. While deviations along the real axis for higher-order roots are still captured by overlapping contours, the situation becomes tenuous for lower-order roots, as the first contour lacks a neighbouring contour on one side to help capture roots. Deviations along the imaginary axis, which become increasingly pronounced at long wavelengths, affect roots of all orders and further complicate their capture. To address these challenges, we introduced the following updates to the algorithm.}


{\XYZ First, we revised the method by replacing the original search variables, $\epsilon_c$ and $\beta$, with $(\alpha_c a)^2$ and $(\beta a)^2$, respectively. This rescaling reduces the sensitivity of the attraction basins of the transcendental equation ($f_m$) to changes in $a$, with the proximity of roots to their singularities now solely determined by the dimensionless product $ka$}. 
In addition, we search for $(\alpha_c a)^2$ rather than $\alpha_c a$, for example, since the latter would result in a duplication of the complex search space. For the implementation of the argument principle method, we provide the explicit expressions of the {\XYZ derivatives of $f_m$} with respect to the updated search variables in~\ref{sec:fiberderiv}.

\begin{figure}[h]
   \begin{center}
   \includegraphics[scale = 0.36]{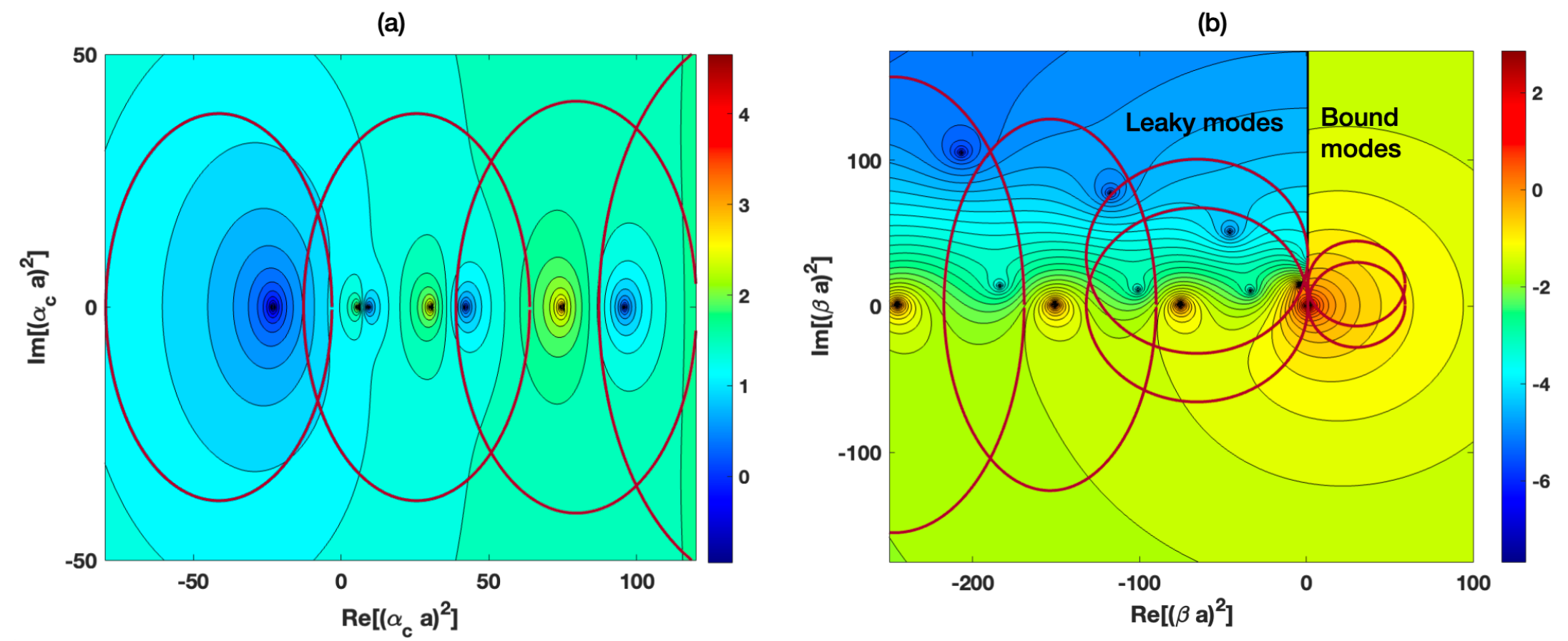}
   \end{center}
   \caption{Examples of the attraction basins and our revised contour selection strategy (magenta) for a circular fiber geometry. Horizontal and vertical axes represent real and imaginary parts of the complex search variable domain, while the color indicates the logarithmic magnitude of the transcendental equation~\eqref{eq:dispersion} ($\log_{10}(|f(z)|)$; see equivalent figure in~\cite{Parry-wire-principal_value} for more details). Thus, the roots correspond to the deep blue regions and the singularities to the red regions. Plot (a) shows the attraction basin of the $(\alpha_c a)^2$ search variable, {\XYZ featuring an enlarged contour (the second from the left) that encloses the first two singularities, as described in Eq.~\eqref{eq:largecontour}. 
   Another contour, the leftmost one defined in Eq.~\eqref{eq:plasmonic}, is included to capture plasmonic roots that may drift further into the negative half} of the complex plane. {\XYZ Plot (b) shows an example of an attraction basin in the $(\beta a)^2$ search variable where a branch point exists at $\beta^2 = k^2 \epsilon_b \mu_b$ with a branch cut selected to extend towards $+i\infty$.}
  }
   \label{fig:contours}
\end{figure}

We also revised our contour strategy for finding the first few roots, via a simple modification that has been tested to succeed even for challenging parameter regimes. In particular, we combine the first two contours into one large contour that covers a much larger region of the search space. Even so, this enlarged contour still does not enclose more than 6 roots, and thus does not present any difficulties for the numerically sensitive Newton's identities. This contour is displayed in Fig.~\ref{fig:contours}. Specifically, we define the center and radius of this enlarged contour to be
\begin{equation}
\begin{aligned}
c_1 &= \frac{3}{8} u_3 + \frac{1}{8} u_2 + \frac{1}{8} u_1 + \frac{3}{8} u_0, \\
r_1 &= \frac{3}{8} u_3 + \frac{1}{8} u_2 - \frac{1}{8} u_1 - \frac{3}{8} u_0.
\end{aligned}
\label{eq:largecontour}
\end{equation}
This equation uses the notation of Eq.~(13) from the original manuscript~\cite{Parry-wire-principal_value}, where $u_k$ is the value of $k$-th singularity, and $u_0 = 2 u_1 - u_2$. 
{\XYZ Additionally, in the context of $(\alpha_c a)^2$ as the search variable, the plasmonic root often drifts further into the negative half of the complex plane. To ensure its capture, we introduce an additional contour positioned along the negative real axis, adjacent to the enlarged first contour. For our purposes, we choose a contour defined by
\begin{equation}
\begin{aligned}
c_p &= c_1 - \dfrac{3}{2}r_1,\\
r_p &= r_1.
\end{aligned}
\label{eq:plasmonic}
\end{equation}}
This additional contour can be further enlarged to cover more of the search space. This does not risk numerical difficulties, since the contour is not expected to enclose many roots, if at all, and is situated in a relatively featureless region of the search space. An example of this contour is displayed in Fig.~\ref{fig:contours}(a). If necessary, the contour can be further enlarged to cover even areas with known roots and singularities by running the additional contour last, and deflating the additional contour of all known roots and singularities. This ensures that the additional contour will not contain more than 5 new roots, which is necessary for the stability of the Newton identities.

{\XYZ To handle deviations of roots along the imaginary axis,} especially at long wavelengths, we employ elliptical contours as roots may wander much farther along the imaginary axis in comparison to the distance between consecutive singularities on the real axis, as shown in Fig.~\ref{fig:contours}(b).
{\XYZ Like the circularly-shaped contours used so far, elliptically-shaped contours are periodic, so the equidistant trapezoidal rule for numerical integration converges exponentially.}

{\XYZ However, elliptical contours, like circular ones, are not particularly effective at preventing failure of our root-search subroutine when the contour is simply shrunk in an attempt to avoid branch points and branch cuts. Thus, we avoid placing elliptical contours adjacent to branch cuts. Instead, we use multiple shifted circular contours to collectively cover the area that would otherwise be enclosed by an elliptical contour.}

To make the root search more robust, we also implement an adaptive scheme to search further deep into the imaginary axis of the complex plane until a root is found, i.e.,
\begin{equation}
\begin{aligned}
    c_k &= c_k \pm l * r_k, \\
    r_k &= r_k.
    \end{aligned}
\end{equation}
Here, $l$ is an integer and its value is incremented until a root is found for every contour.

\subsection{{\XYZ Computation of the leaky ($\beta$) modes}}
\label{sub:Leaky}
{\XYZ In the context of optical waveguide ($\beta$) modes, a transcendental equation such as Eq.~\eqref{eq:dispersion} is traditionally used to describe bound modes, characterized by $\alpha_b$ having a positive imaginary part to ensure field decay away from the core. However, it is also suitable to describe leaky modes, which were not considered in our original implementation~\cite{Parry-wire-principal_value}. Leaky modes exhibit both a radiative nature (i.e., a non-zero real part of $\alpha_b$) and leakage (i.e., a negative imaginary part of $\alpha_b$, leading to exponential growth); the latter reflects the open (i.e., non-Hermitian) nature of the problem.


This treatment of leaky modes contrasts with the more traditional Hermitian description of the problem, which yields a continuum of radiative modes~\cite{Chew-book,Marcuse-book}, i.e., modes for which there is no leakage ($\alpha_b$ is purely real). This Hermitian description requires a slightly more complicated spatial field profile (compared to Eq.~\eqref{eq:EH_field_beta_fiber} in~\ref{sec:fiberfields}), one that also includes incoming waves in the background medium. Consequently, the field profiles and dispersion relations associated with the continuum of radiative modes are inconsistent with the conventional descriptions of bound modes.

An equivalent classification of bound and leaky modes can be made in terms of the eigenvalue $\beta^2$, relative to the light line of the background medium defined by the branch point $\beta^2 = k^2 \epsilon_b \mu_b$ in the complex $\beta^2$ plane. Bound modes lie to the right of this branch point, where $\mathrm{Re}(\beta^2) > k^2 \epsilon_b \mu_b$, while leaky modes lie to the left, where $\mathrm{Re}(\beta^2) < k^2 \epsilon_b \mu_b$. Notably, because $\beta^2$ is generally complex, classification based solely on $\mathrm{Re}(\beta)$ is unreliable.}
{\XYZ This branch point is not only central to mode classification but also critical for the success of our method, as it must be avoided by root-searching contours. In particular, two branch cuts emanate from this branch point, originating separately from the square root function of $\alpha_b$ and the Hankel function. In MATLAB and most other numerical packages, branch cuts are positioned along the positive real axis in the complex domains of $(\beta a)^2$, which typically coincides with the region of the transcendental equation containing the bound modes. To prevent interference with the root-finding contours, it is therefore necessary to reposition these branch cuts.}

{\XYZ The combined effect of the direction of $\alpha_b$ rotation and the choice of sign influences the mode properties, determining whether the modes exhibit the correct radiation and leakage behaviour consistent with our problem formulation.}
%
%
%
{\XYZ Specifically, our original implementation~\cite{Parry-wire-principal_value}, which targeted only bound modes, positioned the branch cuts along the negative imaginary axis and selected the negative square root branch. However, now that we are interested in leaky modes, we realize that this setup produces leaky modes with incoming wave boundary conditions, contradicting our intended outgoing wave formulation. Thus, our updated approach repositions the {\XYZ branch cuts} along the positive imaginary axis and adopts the positive root branch to capture both bound and leaky modes of the transcendental equation with correct mode characterization. This is achieved by defining
\begin{equation}
     \alpha_b =  \sqrt{(k^2 \epsilon_b \mu_b - \beta^2) e^{- i \pi/2}} \, e^{i \pi/4}, \quad H_m(\alpha_b a) = \dfrac{2}{\pi i^{m+1}}K_m(-i\alpha_b a).
\end{equation}
With this configuration, the repositioned branch cuts naturally separate bound and leaky modes, as shown in Fig.~\ref{fig:contours}(b).}



\begin{figure}
    \centering
    \includegraphics[scale=0.25]{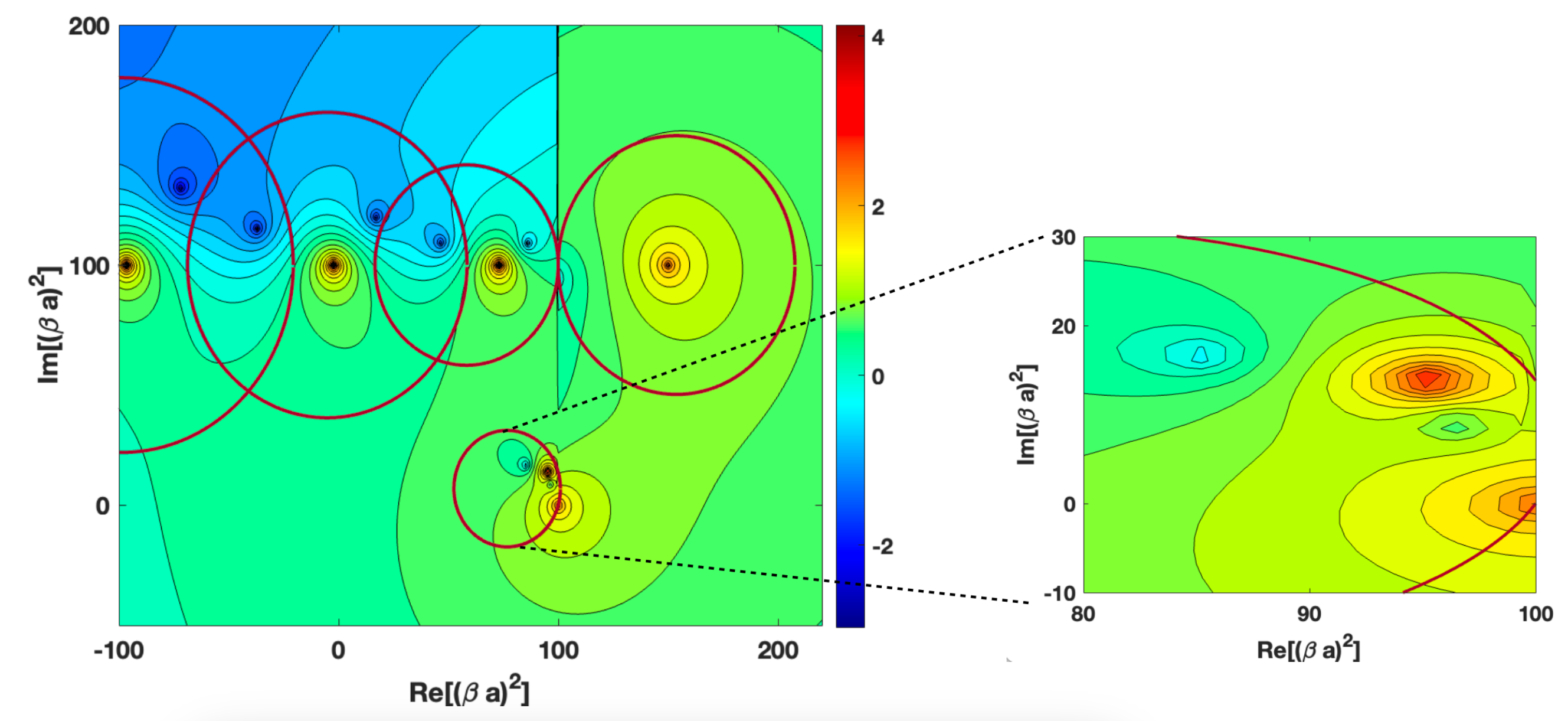}
    \caption{{\XYZ Illustration of the scenario where a singularity arising from the Hankel function lies outside the region covered by the overlapping contours of the Bessel function zeros. An additional contour is introduced around the Hankel singularity, positioned below the overlapping Bessel function contours. The zoomed-in view shows the Hankel singularity and its surrounding roots in very close proximity, clearly demonstrating the algorithm’s effectiveness in resolving roots within narrow attraction basins.}}
    \label{fig:hankel_Contour}
\end{figure}

{\XYZ In addition to the singularities treated in the original implementation, additional singularities may also arise from the complex zeros of the Hankel function, which on the principal branch lies just below the negative real axis and in an eye-shaped region below the origin in the $\alpha_b a$ plane~\cite{Abramowitz-Stegun-book}. During the repositioning of the branch cut along the imaginary axis, singularities arising from the Hankel function vanish in the region containing bound modes. 
{\XYZ The remaining singularities persist in the leaky mode region and tend to cluster tightly near the branch point, unlike the more widely distributed singularities associated with Bessel functions. These leaky-region singularities were previously ignored because the algorithm focused solely on bound modes. To accurately identify roots near the branch cut, it is necessary to include these singularities in the root count evaluation (i.e., for Eq.~(6) in the original manuscript). Furthermore, there are scenarios where the roots in the vicinity of these singularities are not enclosed by the overlapping contours created by Bessel function zeros, as shown in Fig.~\ref{fig:hankel_Contour}. In this update, the algorithm has been extended to include Hankel singularities and, when necessary (as in Fig.~\ref{fig:hankel_Contour}), to introduce dedicated contours around them. Complex zeros of the Hankel function for azimuthal orders up to $m \leq 31$ are now precomputed and incorporated into the root-search module.}}

\begin{figure}[ht!]
   \begin{center}
   \includegraphics[scale = 0.8]{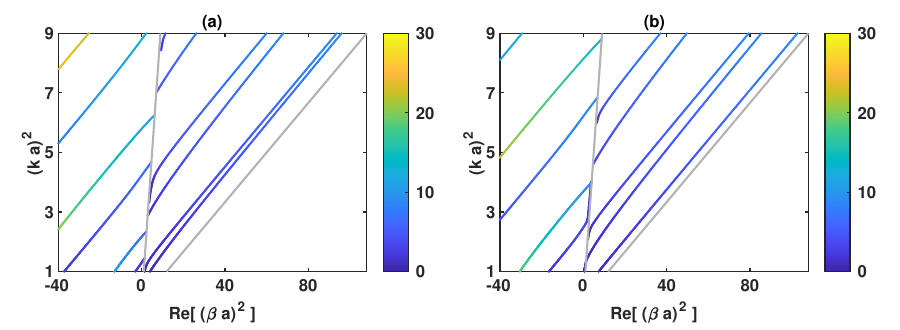}
   \end{center}
   \caption{{\XYZ Dispersion relations produced by our algorithm using complex $(\beta a)^2$ as the search variable for a circular fiber with core permittivity $\epsilon_c = 12 + i$ embedded in a vacuum background medium, where the core radius $a$ is normalized to 1; $\text{Im}[(\beta a)^2]$ is represented by color. Plot (a) shows modes corresponding to the angular order $m = 0$, and plot (b) is for $m = 1$ order. The leftmost grey line in each plot represents the vacuum (background medium) light line, while the other grey line corresponds to  $\beta^2 = 12 k^2$. 
   }}
   \label{fig:dispersion_fiber}
\end{figure}

{\XYZ We demonstrate the computation of the leaky ($\beta$) modes for a numerical example of a non-magnetic circular fiber with core permittivity of $\epsilon_c = 12 + i$ embedded in a vacuum background medium ($\epsilon_b = 1$). In Fig.~\ref{fig:dispersion_fiber}, we present the dispersion characteristics as $\beta^2$ versus $k^2$, instead of the traditional $\beta$ versus $k$ plot, to more clearly distinguish the modes based on their positions relative to the light lines. The plot shows $(ka)^2$ on the vertical axis, $\real[(\beta a)^2]$ on the horizontal axis, and $\imag[(\beta a)^2]$ represented by color.
The light lines of the background medium ($\beta^2 = \epsilon_b k^2$) and of the core medium ($\beta^2 = \epsilon_c k^2$) are shown as grey lines in the plots, where only the real part of the latter is depicted due to the complex nature of $\epsilon_c$. Bound modes lie between the two light lines, while leaky modes are to the left of the background medium light line. We observe that the dispersion curves exhibit a discontinuity as they transition across the background medium light line, which can be attributed to the fact that bound and leaky modes lie on opposite sides of the branch cut (as shown in Fig.~\ref{fig:contours}(b)). A related phenomenon is discussed in~\cite{sven_hidden_Resonances} in the context of hidden multi-sheeted resonances near branch cuts.}


\subsection{Implementation of the algorithm to other geometries}
\label{sub:other_geometries}
{\XYZ In this Section, we extend our root-search algorithm to general asymmetrically bound slab waveguide geometries (including the simplified symmetrical case) and sphere geometries.}

\subsubsection{Slab Waveguide}
\label{sub:slab}
\begin{figure}[ht!]
   \begin{center}
   \includegraphics[scale = 0.25]{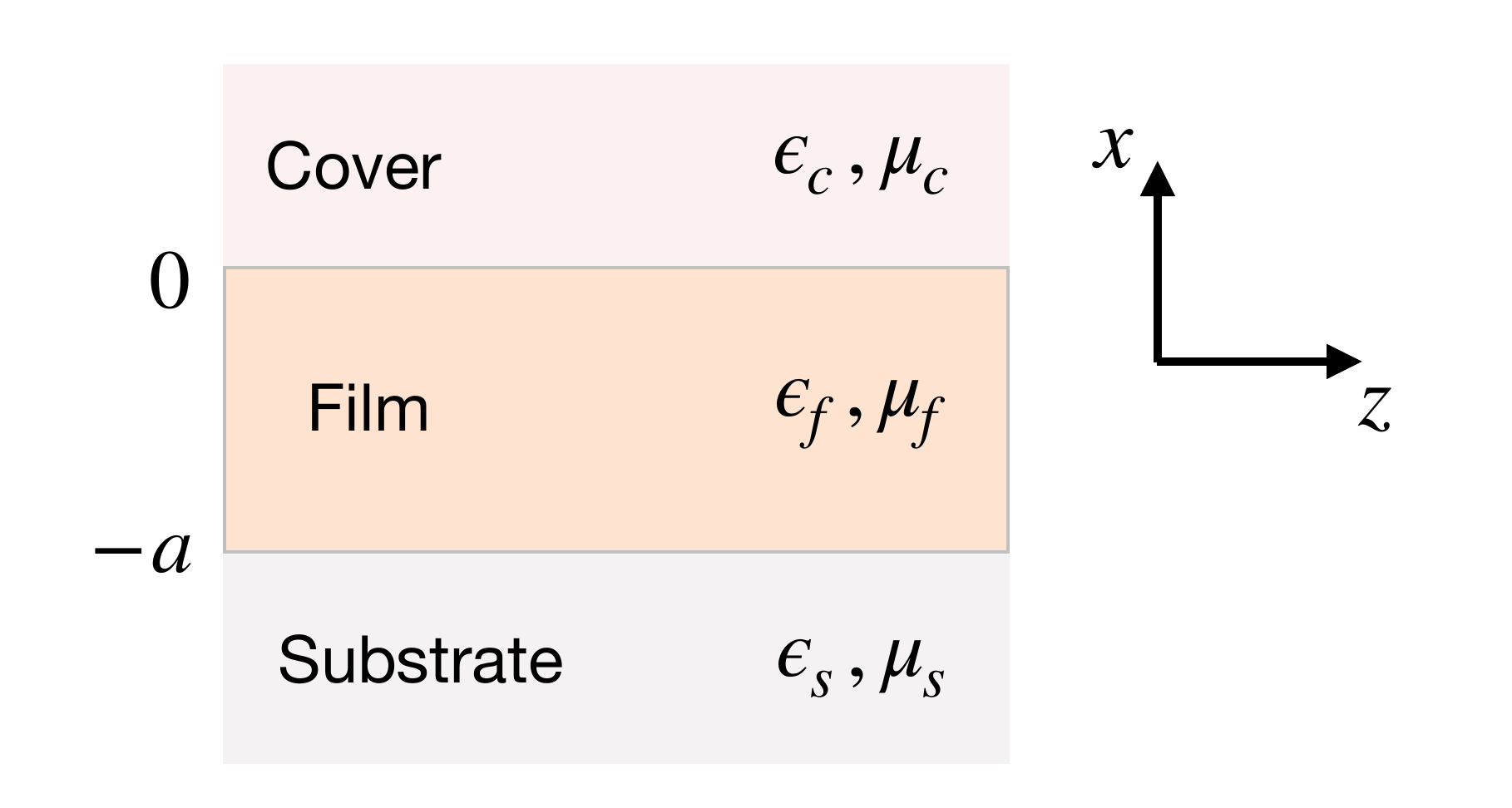}
   \end{center}
   \caption{Schematic of slab waveguide geometry shown in $x-z$ plane, with $y$-axis pointing out of the paper.}
   \label{fig:Slab_geometry}
\end{figure}

{\XYZ We consider an asymmetrically bound slab waveguide geometry consisting of three layers:} a guiding film layer, surrounded by a cover and a substrate, as shown in Fig.~\ref{fig:Slab_geometry}. We denote the permittivities and permeabilities of these layers as $\epsilon_f$ and $\mu_f$, $\epsilon_c$ and $\mu_c$, and $\epsilon_s$ and $\mu_s$, respectively. The guiding layer is infinite in the $y$-$z$ plane and finite along the $x$-axis, with thickness $a$.

In this slab waveguide geometry, both transverse electric (TE) and transverse magnetic (TM) modes are supported. We treat these in turn. First, the transcendental equation of TE modes ($f^\textrm{TE}$) is given by~\cite{Yeh-book}
\begin{equation}
   f^\textrm{TE} \equiv \dfrac{\alpha_f^2 + \alpha_c \alpha_s}{\alpha_f}\tan(\alpha_f a) + i (\alpha_c +  \alpha_s) = 0,
\label{eq:TEdispersion}
\end{equation}
{\XYZ where the transverse propagation constant $\alpha_j$ along the $x$-axis in regions $j \in {c, f, s}$ (cover, film, and substrate, respectively) is defined as
\begin{equation}\label{eq:alpha_gamma}
\alpha_j^2 = k^2 \epsilon_j \mu_j - \beta^2.
\end{equation}}
Here, $k$ is the wavenumber in free space and $\beta$ is the in-plane propagation constant along the $z$-direction. In this transcendental equation, we can set either $(\beta a)^2$ or $(\alpha_f a)^2$ as the eigenvalue or search variable.

Similarly, the transcendental equation for TM modes is given by
\begin{equation}
f^\textrm{TM} \equiv \dfrac{\alpha_f^2 \epsilon_c \epsilon_s + \epsilon_f^2\alpha_c \alpha_s}{\alpha_f \epsilon_f} \tan(\alpha_f a) + i( \epsilon_c \alpha_s + \epsilon_s \alpha_c) = 0.
\label{eq:TMdispersion}
\end{equation}

We obtain the derivatives to the dispersion relations~\eqref{eq:TEdispersion} and~\eqref{eq:TMdispersion} in~\ref{sec:slabderiv}, which are needed for the argument principle method{\XYZ \footnote{As pointed out in our earlier work~\cite[Eq.~(10)]{Parry-wire-principal_value}, the use of the derivative of the dispersion relations is convenient, but not a necessity.}}. Next, we demonstrate the numerical example of our algorithm for the slab waveguide {\XYZ for both search variables: $(\beta a)^2$ and $(\alpha_f a)^2$}.

\begin{figure}[ht!]
   \begin{center}
   \includegraphics[scale = 0.8]{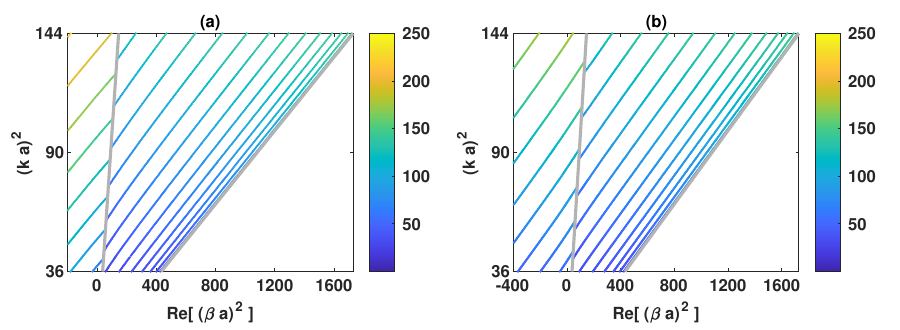}
   \end{center}
   \caption{Dispersion relations of modes of the symmetric slab waveguide, produced by our algorithm with complex $(\beta a)^2$ as the search variable, with $\text{Im}[(\beta a)^2]$ represented by color. Plot (a) is of TE modes, and plot (b) shows TM modes for a guiding film layer of $\epsilon_f = 12 + i$ and vacuum background. The guiding film layer thickness $a$ is normalised to 1. The grey line on the left of each figure indicates the light line of vacuum {\XYZ (i.e., $\beta^2 = k^2$), while the other grey line corresponds to $\beta^2 = 12 k^2$}. The modes between the two light lines are bound modes, i.e., their fields are evanescent in the background (i.e., in the cover and substrate layers). The modes above the first light line are {\XYZ leaky} modes, i.e., their fields grow exponentially in the background.}
   \label{fig:dispersion}
\end{figure}

We demonstrate the numerical example of the algorithm for $(\beta a)^2$ as an eigenvalue for a {\em symmetrical} slab waveguide with guiding layer permittivity of $\epsilon_f = 12 + i$, and vacuum cover and substrate layers. {\XYZ We show dispersion relations ($\beta^2$ vs $k^2$) of TE and TM modes of this slab waveguide in Fig.~\ref{fig:dispersion}, with $\real[(\beta a)^2]$ along the horizontal axis and $\imag[(\beta a)^2]$ represented by color. The light lines of the substrate and cover coincide, given by $\beta^2 = k^2$ (vacuum light line), while the light line of the guiding film layer is $\beta^2 = {Re(\epsilon_f)}k^2$; both are shown as grey lines in the dispersion plots. Since a branch cuts arises in the slab waveguide geometry due to the substrate and cover regions, we observe discontinuities in the dispersion curves across the vacuum light line, similar to what was observed in fibers in Section~\ref{sub:Leaky}.}


\begin{figure}[ht!]
   \begin{center}
   \includegraphics[scale = 0.36]{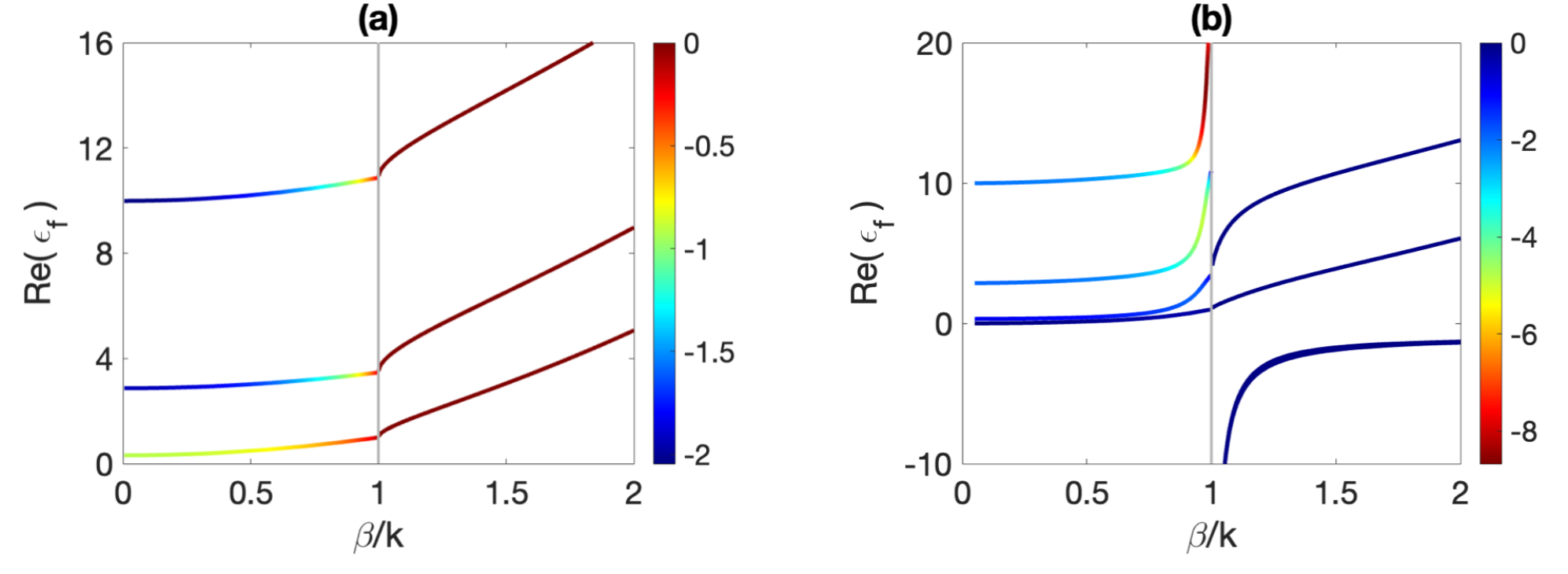}
   \end{center}
   \caption{Dispersion relations of eigenpermittivity modes of symmetrical slab waveguide determined using $(\alpha_f a)^2$ search variable. The results are shown in terms of $\epsilon_f$ of slab, with  $\text{Im}(\epsilon_f)$ indicated by color. Plot (a) is of TE modes, and plot (b) shows TM modes for a wave-number $k = 2$, with a guiding film layer thickness $a$ normalized to 1, and vacuum in both substrate and cover regions. The vertical grey line indicates the light line of vacuum, and the bound modes with only real $\epsilon_f$ exist to the right of the light line, while {\XYZ leaky} modes with complex $\epsilon_f$ exist to the left of the light line.}
\label{fig:dispersion_epsilon}
\end{figure}

{\XYZ Next, we compute the permittivity modes. In this case, $k$ and the permittivities of the substrate and cover are fixed; the propagation constant $\beta$ is the independent variable of the dispersion relation, and the guiding film layer’s permittivity $\epsilon_f$ is the dependent variable. Unlike in~\cite{Parry-wire-principal_value}, here $\epsilon_f$ is determined by treating $(\alpha_f a)^2$ as the search variable. We show the dispersion relations for this case in Fig.~\ref{fig:dispersion_epsilon}. The only relevant light lines are those of the substrate and cover regions, which, in this example, are both vacuum; thus, they coincide, and this vacuum light line ($\beta = k$) appears as a vertical line in the dispersion plot since $k$ is fixed. The bound modes exist to the right of the light line and always have real $\epsilon_f$, while the {\XYZ leaky} modes exist to the left of the light line and have complex $\epsilon_f$. Since branch-cuts don't emerge in the transcendental equations~\eqref{eq:TEdispersion} and~\eqref{eq:TMdispersion} for $(\alpha_f a)^2$ search variable, the dispersion curves do not show any discontinuities across the vacuum light line.}

The explicit expressions for the field profiles of such permittivity modes are given in~\ref{sec:Slab_ModeProfile}.


\newpage

\subsubsection{Spheres}
We now turn to solve the transcendental equations for the modes of a sphere. The relevant parameters are the radius $a$, free-space wavenumber $k$, background permittivity $\epsilon_b$. The modes can have two polarizations: TE, where the component of the electric field along the radial direction is zero, and TM, where the radial magnetic field is zero. Each of these modes is generated by separate dispersion relation equations: 

TM:
\begin{eqnarray}
f_l^{(\textrm{TM})}( x) \equiv \frac{x j_l'(x)}{j_l(x)} - \frac{x_b {h^{(1)}_l}'(x_b)}{h^{(1)}_l(x_b)} = 0,
\label{eq:TM_roots_search}
\end{eqnarray}

TE:
\begin{eqnarray}
f_l^{(TE)}( x) \equiv \frac{1}{x^2} \left(1 + \frac{x j_l'(x)}{j_l(x)}\right) - \frac{1}{k_b^2} \left(1 + \frac{x_b {h^{(1)}_l}'(x_b)}{h^{(1)}_l(x_b)}\right) = 0, \nonumber \\
\label{eq:TE_roots_search}
\end{eqnarray}
Here, $x = \sqrt{\epsilon} k a$, $j_l$ and $h^{(1)}_l$ are the spherical Bessel and spherical Hankel functions of the first kind of order $l$, and the prime denotes differentiation with respect to the entire argument. {\XYZ For a given polar order $l$ and polarization ($F = TM$ or $TE$), we get an infinite sequence of roots which are enumerated by the radial order $n$, as mentioned in Appendix C. no additional information from the angular dependence affect the eigenvalues.  

From repeated numerical experimentation, we observe that, as found for wire modes in~\cite{Parry-wire-principal_value}, all the photonic roots in the $f_l^{(\textrm{TM})}$ case are relatively close to their neighboring singularities (spherical Bessel function zeroes). Thus, the locations of the singularities, known analytically, provide guideposts for the locations of the roots and the center of the contours. However, the TE polarization has a special root that can be far from the singularities (for small spheres ($k a \ll 1$), it is a plasmonic root) or close to other root. In case it is beside other roots may cause the surrounding roots to move away or toward the closer singularity and could lead to
troublesome features such as closely spaced roots, sharp resonances, and the presence of singularities near the roots, all of which can hinder the reliability and completeness of root-finding processes of $f_l^{(TE)}$. The special TE root has the largest imaginary absolute value. Only after finding all the other roots, we use a large contour to find it (this contour may contain more than two roots). }


{\XYZ In both the TM and TE cases, we choose to use circular contours, which are the simplest to implement. In many cases overlapping circles help us to cover the same root twice and give the method another opportunity to converge to the root using another initial guess. In some rare cases, we can use a smaller contour around the previous initial guess for the Newton-Raphson method that did not converge to a root.

In the $f_l^{(TE)}$ case, we observe that the roots of $f_l^{(TE)}$ are close to the roots of $f_l^{(TE)}$, therefore we locate the center of contour between these roots. We can easily find the roots of $f_l^{(TE)}$ using the Newton-Raphson method with initial guess close to the singularities of $f_l^{(TE)}$.}


{\XYZ Once again, to apply the argument principle method, we require the derivatives of~\eqref{eq:TM_roots_search} and~\eqref{eq:TE_roots_search}. Because of their similarity to the fiber dispersion relation~\eqref{eq:dispersion} and because derivatives spherical and cylindrical Bessel functions obey the same identities, we omit the detailed derivations, and refer the reader to the derivations of~\ref{sec:fiberderiv}. As for the geometries studied above the algorithm can be applied for either frequency or permittivity modes.}


\begin{figure}[ht!]
   \begin{center}
   \includegraphics[scale = 0.2]{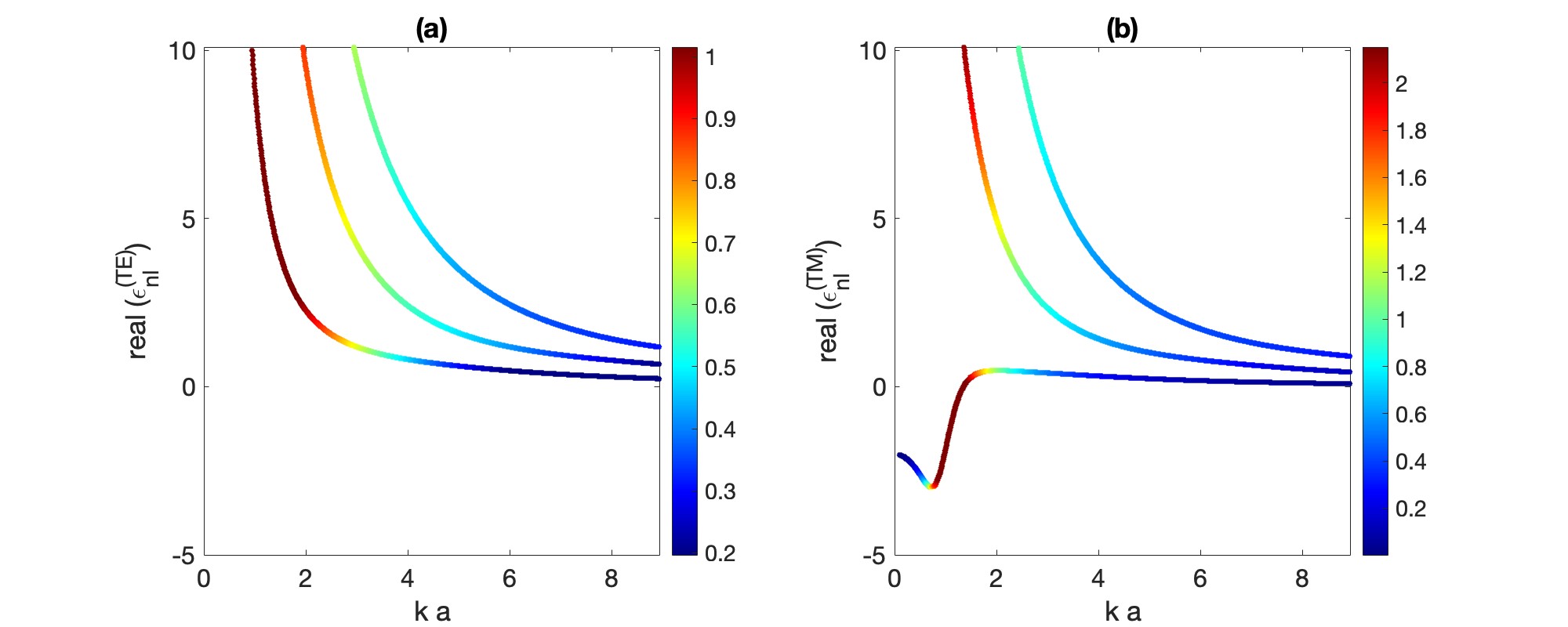}
   \end{center}
   \caption{Dispersion relations of eigenpermittivity modes of a sphere determined using $(k a)$ search variable (where $k$ is a constant). The y axis represents the real value of $\epsilon_{nl}$, while $|(\text{Im}(\epsilon_{nl})|$ is indicated by color. Each line represents different $n$ index ($n = 1..3$), where $l$ is constant ($l = 1$). Plot (a) is of TE modes and plot (b) shows TM modes in a vacuum background.} 
\label{fig:dispersion_epsilon_sphere}
\end{figure}


{\XYZ Fig.~\ref{fig:dispersion_epsilon_sphere} shows the behavior of the eigenpermittivities of a sphere as a function of $k a$ (where $k$ is a constant).  Each line represents different $n$ index, where $l$ is constant ($l = 1$).
One can see that the TE eigenvalues are all positive, their real and imaginary parts decrease monotonically with growing radius. The behaviour of the TM eigenvalues is similar, with the exception of the $n = 1$ case for which the eigenvalue becomes negative (i.e., plasmonic, $\epsilon_{nl}^{TM} \approx -2$) and purely real for small radii ($k a < 0.1$).}


\section{Outlook}
We have described various improvements and extensions of our argument principle root-search applied to various generic optical structures. We specifically demonstrated propagation constant ($\beta$) and permittivity modes, but our algorithm can also be applied to frequency modes~\cite{lalanne_2018_LPR} (aka resonant states~\cite{Muljarov_2010}). In addition to the usefulness of the algorithms detailed above per se, they also provide a convenient starting point for perturbation techniques (in the context of resonant state expansions~\cite{Muljarov_2010} or re-expansion~\cite{Parry-Egor-re-expansion,Kossowski-Chen,Parry-GALM}). They can also be extended to simple geometries with more sophisticated physics, such as for systems with anisotropic, nonlocal~\cite{GdA_nonlocal_2008,DTU_nonlocal,Mortensen_nonlocality_quantum_size,TO_vdw_PNAS,Antonio_kissing_cyls_nonlocal_PRL,nonlocal_IMI_modes} or non-reciprocal response. Finally, our algorithm can also be applied for the computation of electron wavefunctions in an arbitrary potential in atomic calculations~\cite{Moiseyev-book}, {\XYZ offering an alternative to the Riesz-projection contour integral method~\cite{sven_riesz_nonlinear}, which requires operator discretization and results in a nonlinear eigenvalue problem}. All these potential directions would be explored in future research.

\section{Acknowledgments. } S. R. and Y. S. were partially funded by Israel Science Foundation (ISF) grant no. 340/2020 and DFG (Deutsche Forschungsgemeinschaft – Middle East Collaboration) grant no. WE 5815/5-1.

\appendix

\section{Circular fiber geometry}
{\XYZ In this Appendix, we present the analytical derivatives of the transcendental equation for circular fibers, which are required for our root-search algorithm, along with the corresponding field profiles associated with the roots.}

\subsection{{\XYZ Derivatives of transcendental equation}}
\label{sec:fiberderiv}
{\XYZ We evaluate the derivatives used by the argument principle and Newton's methods for both $(\alpha_c a)^2$ and $(\beta a)^2$ search variables.
For notational simplicity, we introduce new symbols for the transcendental equation~\eqref{eq:dispersion}},
\begin{align}
R^J_m &= \frac{J'_m(\alpha_c a)}{\alpha_c a J_m(\alpha_c a)}, & R^H_m &= \frac{H'_m(\alpha_b a)}{\alpha_b a H_m(\alpha_b a)}.
\end{align}
Thus, the {\XYZ transcendental equation~\eqref{eq:dispersion}} can be rewritten as
\begin{equation}
f_m = (\epsilon_c R^J_m - \epsilon_b R^H_m) (\mu_c R^J_m - \mu_b R^H_m)  - \left(\frac{m\beta}{k}\right)^2 \left(\frac{1}{(\alpha_c a)^2} - \frac{1}{(\alpha_b a)^2}\right)^2.
\label{eq:disprelr}
\end{equation}
{\XYZ We first compute the derivative of $f_m$ with respect to $(\alpha_c a)^2$, given by
\begin{equation}
\begin{split}
\dfrac{\partial f_m}{\partial (\alpha_c a)^2} =&  \left(\frac{\partial \epsilon_c}{\partial (\alpha_c a)^2} R^J_m + \epsilon_c \frac{\partial R^J_m}{\partial (\alpha_c a)^2} \right) (\mu_c R^J_m - \mu_b R^H_m)\\
& + (\epsilon_c R^J_m - \epsilon_b R^H_m) \mu_c \frac{\partial R^J_m}{\partial (\alpha_c a)^2}  + \frac{2m^2\beta^2}{(\alpha_c a)^4 k^2} \left(\frac{1}{(\alpha_c a)^2} - \frac{1}{(\alpha_b a)^2}\right),
\end{split}
\end{equation}
where
\begin{equation}
\frac{\partial \epsilon_c}{\partial (\alpha_c a)^2} = \frac{1}{(ka)^2 \mu_c}.
\end{equation}
This expression follows from the relation $\alpha_c^2 + \beta^2 = k^2\mu_c \epsilon_c$. In addition,
\begin{equation}
\frac{\partial R^J_m}{\partial (\alpha_c a)^2}  = -\dfrac{1}{2 (\alpha_c a)^2}\left[1 + 2 R^J_m  - \dfrac{J_{m+1}(\alpha_c a) J_{m-1}(\alpha_c a)}{J_m^2(\alpha_c a)}\right],
\label{eq:drj}
\end{equation}
which is derived using the Bessel function's defining differential equation}.

Next, we obtain the derivative of $f_m$ with respect to $(\beta a)^2$, yielding
\begin{equation}
\begin{split}
\dfrac{\partial f_m}{\partial (\beta a)^2} =&
\left(\epsilon_b \frac{\partial R^H_m}{\partial (\alpha_b a)^2} - \epsilon_c \frac{\partial R^J_m}{\partial (\alpha_c a)^2} \right) (\mu_c R^J_m - \mu_b R^H_m) \\ & +\left(\mu_b\frac{\partial R^H_m}{\partial (\alpha_b a)^2} - \mu_c\frac{\partial R^J_m}{\partial (\alpha_c a)^2} \right) (\epsilon_c R^J_m - \epsilon_b R^H_m) \\
& - \left(\frac{m}{ka}\right)^2\left(\frac{1}{(\alpha_c a)^2} - \frac{1}{(\alpha_b a)^2}\right)^2\left(1+2 \left(\frac{\beta a}{\alpha_c a}\right)^2 + 2 \left(\frac{\beta a}{\alpha_b a}\right)^2 \right),
\end{split}
\end{equation}
{\XYZ where the chain rule was used along with the identity
\begin{equation}
\frac{\partial (\alpha_{j} \, a)^2}{\partial (\beta a)^2} = - 1, \qquad j \in \{ c,b\}.
\end{equation}
Since the Hankel functions satisfy identities analogous to those of the Bessel functions, the expression of derivative $\partial R^H_m/\partial (\alpha_b a)^2$ can be obtained by substituting $\alpha_c \rightarrow \alpha_b$, $ R^J_m \rightarrow R^H_m $ and $J_m(\alpha_c a) \rightarrow H_m(\alpha_b a)$ in Eq.~\eqref{eq:drj}.}

\subsection{Fiber mode field profiles}
\label{sec:fiberfields}
{\XYZ In this Appendix, we provide explicit expressions for the eigenmode fields of a step-index circular fiber, along with the normalization constants for both $\epsilon_c$ and $\beta$ modes. The mode fields are expressed in the cylindrical coordinate system $(\rho,\theta,z)$, and they take a separable form
\begin{equation}
\bv{E}(\rho,\theta,z) = A \, \bv{e}(\rho) e^{i m \theta} e^{i \beta z}, \quad     \bv{H}(\rho,\theta,z) = A \, \bv{h}(\rho) e^{i m \theta} e^{i \beta z}. \label{eq:fiber_mode_eqs}
\end{equation}
Here, $A$ denotes the normalization constant (also referred to as the mode amplitude), and the vectors $\bv{e}(\rho)$ and $\bv{h}(\rho)$ represent the radial variation of the electric and magnetic fields, respectively. 
Their axial components, $e_z(\rho)$ and $h_z(\rho)$, are sufficient to reconstruct the full vectors, as the transverse ($\rho$ and $\theta$) components can be obtained via
\begin{equation}
\begin{split}
e_r (\rho) = \dfrac{i \beta}{\alpha_j^2} \dfrac{\partial e_z}{\partial \rho} - \dfrac{m k \mu_j}{\alpha_j^2 \rho} h_z, \qquad
e_\theta (\rho) =  \dfrac{-m \beta}{\alpha_j^2 \rho}  e_z - \dfrac{i k \mu_j}{\alpha_j^2} \dfrac{\partial h_z}{\partial \rho}, \\
h_r (\rho) = \dfrac{i \beta}{\alpha_j^2} \dfrac{\partial h_z}{\partial \rho} + \dfrac{m k \epsilon_j}{\alpha_j^2 \rho} e_z,  \qquad
h_\theta (\rho) =  \dfrac{-m \beta}{\alpha_j^2 \rho}  h_z + \dfrac{i k \epsilon_j}{\alpha_j^2} \dfrac{\partial e_z}{\partial \rho},
\end{split}
\label{eq:fiber_transverse_eqs}
\end{equation}
where $j \in \{c, b\}$ denotes the region (core or background, respectively), and the parameters $\alpha_j$, $\mu_j$, and $\epsilon_j$ take values specific to that region. In the above expressions, the mode indices (i.e., the subscripts $m$ and $n$ which denote the azimuthal and radial orders) are omitted to maintain generality across both $\epsilon_c$ and $\beta$ modes. The indices are introduced in the following subsections for quantities that depend explicitly on the corresponding eigenvalue.

\subsubsection{$\epsilon_c$ modes}
We now introduce the mode indices $mn$ to the eigenvalue (i.e., $\epsilon_{c,mn}$) and its dependent quantities: $\alpha_{c,mn}$, normalization constant $A_{mn}$, full fields ($\bv{E}_{mn}$, $\bv{H}_{mn}$), and radial field profiles ($\bv{e}_{mn}$, $\bv{h}_{mn}$). Thus, Eq.~\eqref{eq:fiber_mode_eqs} takes the form
\begin{equation}
\begin{split}
    \bv{E}_{mn}(\rho,\theta,z) &= A_{mn} \, \bv{e}_{mn}(\rho) e^{i m \theta} e^{i \beta z}, \\     \bv{H}_{mn}(\rho,\theta,z) &= A_{mn} \, \bv{h}_{mn}(\rho) e^{i m \theta} e^{i \beta z}.
    \end{split}
\end{equation}
The expression of the axial ($z$) components of $\bv{e}_{mn}$, $\bv{h}_{mn}$ are given by
\begin{equation}
\begin{split}
    e_{z,mn} &= f_{mn}(\rho) , \\
    h_{z,mn} &= H_{0,{mn}} f_{mn}(\rho),
\end{split}
\end{equation}
where
\begin{equation}
   f_{mn}(\rho) =   \begin{cases} \dfrac{J_m(\alpha_{c,{mn}} \rho)}{J_m(\alpha_{c,{mn}} a)}    & \text{for} \quad r\leq a \quad (\text{core}), \\ \\
\dfrac{H_m(\alpha_b \rho)}{H_m(\alpha_b a)}   & \text{for} \quad r > a \quad (\text{background}),
\end{cases}
\end{equation}
and
\begin{equation}
    H_{0,{mn}} = \left(\frac{1}{\alpha_{c,{mn}}^2}-\frac{1}{\alpha_b^2}\right) \dfrac{m \beta}{ka \left[ \dfrac{J_m'(\alpha_{c,{mn}} a)}{\alpha_{c,{mn}}  J_m(\alpha_{c,{mn}} a)} - \dfrac{H_m'(\alpha_b a)}{\alpha_b  H_m(\alpha_b a)}\right]}.
\end{equation}
The normalization constants $A_{mn}$ for $\epsilon_{c,{mn}}$ modes are given by
\begin{equation}
\begin{split}
    A_{mn} &= \sqrt{\dfrac{1}{2\pi \bigintsss_{0}^{a} \left(-e_{r,{mn}}^2 + e_{\theta,{mn}}^2 + e_{z,{mn}}^2\right) r \, dr}},
\end{split}
\end{equation}
where this expression follows from the bi-orthogonality condition of the eigenpermittivity modes~\cite{Bergman_Stroud_PRB_1980,GENOME}. The transverse components $e_{r,{mn}}$ and $e_{\theta,{mn}}$ are obtained by substituting the axial components $e_{z,{mn}}$ and $h_{z,{mn}}$ into the expressions given in~\eqref{eq:fiber_transverse_eqs}.

\subsubsection{$\beta$ modes}
Similarly, we introduce the mode indices $mn$ for the eigenvalue $\beta_{mn}$ and its associated quantities: $\alpha_{c,{mn}}$, $\alpha_{b,{mn}}$, $A_{mn}$, fields ($\bv{E}_{mn}$, $\bv{H}_{mn}$), and radial field profiles ($\bv{e}_{mn}$, $\bv{h}_{mn}$). Accordingly, Eq.~\eqref{eq:fiber_mode_eqs} becomes
\begin{equation}
\begin{split}
    \bv{E}_{mn}(\rho,\theta,z) &= A_{mn} \, \bv{e}_{mn}(\rho) e^{i m \theta} e^{i \beta_{mn} z},  \\   \bv{H}_{mn}(\rho,\theta,z) &= A_{mn} \, \bv{h}_{mn}(\rho) e^{i m \theta} e^{i \beta_{mn} z}.
    \end{split}
    \label{eq:EH_field_beta_fiber}
\end{equation}
The axial components of $\bv{e}_{mn}$, $\bv{h}_{mn}$ are expressed as
\begin{equation}
\begin{split}
    e_{z,{mn}} &=  f_{mn}(\rho) , \\
    h_{z,{mn}} &=  H_{0,{mn}} f_{mn}(\rho)
\end{split}
\end{equation}
where
\begin{equation}
   f_{mn}(\rho) =   \begin{cases} \dfrac{J_m(\alpha_{c,{mn}} \rho)}{J_m(\alpha_{c,{mn}} a)}    & \text{for} \quad r\leq a \quad (\text{core}),  \\ \\
\dfrac{H_m(\alpha_{b,{mn}} \rho)}{H_m(\alpha_{b,{mn}} a)}   & \text{for} \quad r > a \quad (\text{background}),
\end{cases}
\end{equation}
and
\begin{equation}
    H_{0,{mn}} = \left(\frac{1}{\alpha_{c,{mn}}^2}-\frac{1}{\alpha_{b,{mn}}^2}\right) \dfrac{m \beta_{mn}}{k a\left( \dfrac{J_m'(\alpha_{c,{mn}} a)}{\alpha_{c,mn} J_m(\alpha_{c,{mn}} a)} - \dfrac{H_m'(\alpha_{b,{mn}} a)}{\alpha_{b,{mn}} H_m(\alpha_{b,{mn}} a)}\right)}.
\end{equation}
The normalization constant $A_{mn}$ for the $\beta_{mn}$ mode is computed using the expression
\begin{equation}
A_{mn} = \sqrt{\dfrac{1}{ -2\pi i L_{mn} + 4\pi \bigintsss_{0}^a \left(e_{r,{mn}} h_{\theta,{mn}} + e_{\theta,{mn}} h_{r,{mn}}\right) r \, dr}},
\label{eq:Beta_fiber_amp}
\end{equation}
where
\begin{equation}
    L_{mn} =  \left[h_{z,{mn}} \dfrac{\partial e_{\theta,{mn}}}{\partial \beta_{mn}} - e_{\theta,{mn}} \dfrac{\partial h_{z,{mn}}}{\partial \beta_{mn}} + e_{z,{mn}} \dfrac{\partial h_{\theta,{mn}}}{\partial \beta_{mn}} - h_{\theta,{mn}} \dfrac{\partial e_{z,{mn}}}{\partial \beta_{mn}} \right]_{r=a^+}.
\end{equation}
Here, the term $L$ is evaluated at $r = a^+$, i.e., on the background side of the core–background interface. This expression for $A_{mn}$ follows from the bi-orthogonality condition of $\beta_{mn}$ modes and ensures that it remains non-zero even for exponentially diverging leaky modes. The transverse components ($e_{r,{mn}}, e_{\theta,{mn}}$) and ($h_{r,{mn}}, h_{\theta,{mn}}$), required to evaluate~\eqref{eq:Beta_fiber_amp}, are obtained by substituting the axial components $e_{z,{mn}}$ and $h_{z,{mn}}$ into the expressions given in~\eqref{eq:fiber_transverse_eqs}.}


\section{Slab waveguide}
{\XYZ In this Appendix, we present the analytical derivatives of the transcendental equations associated with the general asymmetrically bound slab waveguide geometry, which are used in our root-search algorithm, along with the corresponding field profiles of the roots.

\subsection{Derivatives of transcendental equations}
\label{sec:slabderiv}
We derive the derivatives of the transcendental equations for TE and TM modes in slab waveguide geometry with respect to the search variables $(\alpha_f a)^2$ and $(\beta a)^2$.

\subsubsection{TE modes}
For notational simplicity, we define
\begin{equation}
  R^\textrm{TE} = \dfrac{(\alpha_f a)^2+(\alpha_c a)(\alpha_s a)}{\alpha_f a},
\end{equation}
so that the TE modes transcendental equation~\eqref{eq:TEdispersion} becomes
\begin{equation}
 f^\textrm{TE} = R^\textrm{TE} \tan(\alpha_f a)+i(\alpha_c a+\alpha_s a).
  \label{eq:disp_RTE}
\end{equation}
The derivative of $f^\textrm{TE}$ with respect to $(\alpha_f a)^2$ is
\begin{equation}
\dfrac{\partial f^\textrm{TE}}{\partial (\alpha_f a)^2} = \bigg(\dfrac{\partial R^\textrm{TE}}{\partial (\alpha_f a)^2} \tan(\alpha_f a) + \dfrac{R^\textrm{TE}}{2 \alpha_f a} \sec^2(\alpha_f a)\bigg),
\label{eq:slab_der_alphac}
\end{equation}
where
\begin{equation}
\dfrac{\partial R^\textrm{TE}}{\partial (\alpha_f a)^2} = \dfrac{(\alpha_f a)^2 - (\alpha_c a)(\alpha_s a)}{2 (\alpha_f a)^3}. \label{eq:R_der_alpha_f}
\end{equation}

Next, we obtain the derivative of $f^\textrm{TE}$ with respect to $(\beta a)^2$ as
\begin{equation}
\begin{split}
\dfrac{\partial f^\textrm{TE}}{\partial (\beta a)^2}  = \bigg( -\dfrac{\partial R^\textrm{TE}}{\partial (\alpha_f a)^2}- \dfrac{\partial R^\textrm{TE}}{\partial (\alpha_c a)^2} -\dfrac{\partial R^\textrm{TE}}{\partial (\alpha_s a)^2}\bigg) \tan(\alpha_f a) \\
- \dfrac{R^\textrm{TE}}{2 \alpha_f a}  \sec^2(\alpha_f a) + \dfrac{ i}{2 \alpha_c a} + \dfrac{i}{2 \alpha_s a},
\end{split}
\label{eq:slab_der_beta}
\end{equation}
where the chain rule was used along with the identity
\begin{equation}
\dfrac{\partial (\alpha_{j} \, a)^2}{\partial(\beta a)^2} = -1, \qquad j \in \{c,f,s\}.
\end{equation}
In addition to~\eqref{eq:R_der_alpha_f}, the remaining derivatives of $R^\textrm{TE}$ used in~\eqref{eq:slab_der_beta} are
\begin{equation}
\dfrac{\partial R^\textrm{TE}}{\partial (\alpha_c a)^2} = \dfrac{\alpha_s a}{2(\alpha_f a) (\alpha_c a)},  \qquad \dfrac{\partial R^\textrm{TE}}{\partial (\alpha_s a)^2} =  \dfrac{\alpha_c a}{2(\alpha_f a) (\alpha_s a)}.
\end{equation}


\subsubsection{TM modes}
Similarly to the TE case, we define
\begin{equation}
R^\textrm{TM} = \dfrac{(\alpha_f a)^2 \epsilon_c \epsilon_s +\epsilon_f^2(\alpha_c a)(\alpha_s a)}{(\alpha_f a)\epsilon_f}.
\end{equation}
so that the TM modes transcendental equation~\eqref{eq:TMdispersion} becomes
\begin{equation}
 f^\textrm{TM} = R^\textrm{TM} \tan(\alpha_f a) + i(\epsilon_c \alpha_s + \epsilon_s \alpha_c).
  \label{eq:disp_RTM}
\end{equation}
The derivative of $f^\textrm{TM}$ with respect to $(\alpha_f a)^2$ is
\begin{equation}
\dfrac{\partial f^\textrm{TM}}{(\alpha_f a)^2}  = \bigg( \dfrac{\partial R^\textrm{TM}}{\partial (\alpha_f a)^2} \tan(\alpha_fa) + \dfrac{R^\textrm{TM}}{2 \alpha_f a} \sec^2(\alpha_f a)\bigg),
\end{equation}
where
\begin{equation}
\begin{split}
\dfrac{\partial R^\textrm{TM}}{\partial (\alpha_f a)^2} = \dfrac{1}{(\alpha_f a) \epsilon_f} \Bigg[\epsilon_c \epsilon_s + (\alpha_c a)(\alpha_s a)\dfrac{\partial \epsilon_f^2} {\partial(\alpha_f a)^2}\Bigg] - \dfrac{R^\textrm{TM}}{2 (\alpha_f a)^2} \\ -  \dfrac{R^\textrm{TM}}{\epsilon_f}\dfrac{\partial \epsilon_f}{\partial (\alpha_f a)^2}.
\end{split}
\label{eq:R_der_alpha_f_TM}
\end{equation}
From the relation $\alpha_f^2 + \beta^2 = k^2 \mu_f \epsilon_f$, it follows
\begin{equation}
\dfrac{\partial \epsilon_f}{\partial (\alpha_f a)^2} = \dfrac{1}{(k a)^2 \mu_f}, \qquad \dfrac{\partial \epsilon_f^2}{\partial (\alpha_f a)^2} = \dfrac{2 \epsilon_f}{(k a)^2 \mu_f}.
\end{equation}

Next, we obtain the derivative of $f^\textrm{TM}$ with respect to $(\beta a)^2$ as
\begin{equation}
\begin{split}
    \dfrac{\partial  f^\textrm{TM}}{(\beta a)^2} = \bigg( -\dfrac{\partial R^\textrm{TM}}{\partial (\alpha_f a)^2}- \dfrac{\partial R^\textrm{TM}}{\partial (\alpha_c a)^2} - \dfrac{\partial R^\textrm{TM}}{\partial (\alpha_s a)^2}\bigg) \tan(\alpha_f a) \\
    - \dfrac{R^\textrm{TM}}{2 \alpha_f a} \sec^2(\alpha_f a) + \dfrac{i\epsilon_c}{2 \alpha_s a} + \dfrac{i\epsilon_s}{2 \alpha_c a},
\end{split}
\label{eq:slab_der_beta_TM}
\end{equation}
where the chain rule was used along with the identity
\begin{equation}
\dfrac{\partial (\alpha_{j} \, a)^2}{\partial(\beta a)^2} = -1, \qquad j \in \{c,f,s\}.
\end{equation}
In addition to~\eqref{eq:R_der_alpha_f_TM}, the remaining derivatives of $R^\textrm{TM}$ used in~\eqref{eq:slab_der_beta_TM} are
\begin{equation}
\dfrac{\partial R^\textrm{TM}}{\partial (\alpha_c a)^2} = \dfrac{\epsilon_f (\alpha_s a)}{2 (\alpha_f a)\alpha_c a}, \qquad
\dfrac{\partial R^\textrm{TM}}{\partial (\gamma_s a)^2} = \dfrac{\epsilon_f (\alpha_c a)}{2 (\alpha_f a)\alpha_s a}.
\end{equation}
}

\subsection{Mode field profiles}
\label{sec:Slab_ModeProfile}
{\XYZ In this Appendix, we define the explicit expressions for the mode field profiles of slab waveguide geometry and their corresponding normalization constants for eigenpermittivity ($\epsilon_f$) and in-plane propagation constant ($\beta$) modes.

\subsubsection{$\epsilon_f$ modes}
For TE modes, the electric field is polarized along the $y-$direction (transverse to the plane of incidence, $x-z$), and its modal profile is expressed as
\begin{equation}
    \bv{E}_n(x,z) = \hat{y} \, A_n^{\textrm{TE}} \, u_n(x) \, e^{i\beta z},
\end{equation}
where $u_n(x)$ represents the transverse spatial variation of the mode, with its piecewise form given by
\begin{equation}
u_n(x) = \begin{cases}
     \cos(\phi_n^\textrm{TE}) \: e^{i\alpha_c x}, & \text{for}\ x > 0 \; (\text{cover}), \\
    \cos\left(\alpha_{f,n} x + \phi_n^\textrm{TE}\right), &  \text{for}\ -a \leq x \leq 0 \;(\text{film}), \\
    \cos\left(\alpha_{f,n} a - \phi_n^\textrm{TE}\right) \: e^{-i \alpha_{s} (x + a)}, & \text{for}\ x < -a \; (\text{substrate}),
    \end{cases}
\end{equation}
and the phase shift $\phi_n^\textrm{TE}$ is given by
\begin{equation}
\phi_n^\textrm{TE} = \tan^{-1}\left(\dfrac{-i \alpha_c}{\alpha_{f,n}}\right).
\end{equation}
The index $n$ represents the $n^{th}$ eigenpermittivity mode, and the normalization constant $A_n^\textrm{TE}$ is given by
\begin{equation}
\begin{split}
    A_n^\textrm{TE} &= \dfrac{1}{\sqrt{\int_{-a}^0 u_n^2(x) dx}} = \dfrac{1}{\sqrt{\dfrac{a}{2} + \dfrac{1}{2 \alpha_{f,n}} \sin\left(\alpha_{f,n}a\right) \cos\left(\alpha_{f,n} a - 2 \phi_n^\textrm{TE}\right)}}.
\end{split}
\end{equation}
This normalization expression follows from the bi-orthogonality condition of the eigenpermittivity modes.

For TM modes, since the electric field has nonzero components along both the
$x$ and $z$ directions, it is standard to represent the mode profile in terms of the
$y-$polarized magnetic field, i.e.,
\begin{equation}
\bv{H}_n(x,z) = \hat{y}\, A_n^\textrm{TM}\, v_n(x)\, e^{i \beta z}
\end{equation}
where the transverse spatial variation of the magnetic field, $v_n(x)$, is given by
\begin{equation}
v_n(x) = \begin{cases}
    \cos(\phi_n^\textrm{TM})\, e^{i\alpha_c x}  , & \text{for}\ x > 0, \\
    \cos\left(\alpha_{f,n} x + \phi_n^\textrm{TM} \right), &  \text{for}\ -a \leq x \leq 0, \\
    \cos\left(\alpha_{f,n} a - \phi_n^\textrm{TM} \right) \; e^{-i \alpha_s (x + a)} , & \text{for}\ x < -a,\\
    \end{cases}
\end{equation}
and the phase shift $\phi_n^\textrm{TM}$ is defined as
\begin{equation}
\phi_n^\textrm{TM} = \tan^{-1}\left( \dfrac{-i \epsilon_{f,n}\, \alpha_c}{\epsilon_c\, \alpha_{f,n}}\right).
\end{equation}
Using this magnetic field representation, the corresponding electric field vector for TM modes can be written as
\begin{equation}
\bf{E}_n (x,z) = A_n^\textrm{TM} \bf{e}_n(x) e^{i \beta z},
\end{equation}
where the components of the transverse spatial variation vector, $\bf{e}_n(x)$, in regions $j \in \{c,f,s\}$ is related to $v_n(x)$ as follows:
\begin{equation}
e_{x,n} =  \dfrac{\beta}{k \epsilon_j} v_n(x) ,\quad e_{z,n} = \dfrac{i}{k \epsilon_j} \dfrac{\partial v_n(x)}{\partial x}.
\end{equation}
The normalization constant $A_n^\textrm{TM}$ is then determined based on the biorthogonality condition for the electric field, and is given by
\begin{equation}
\begin{split}
     A_n^\textrm{TM} &= \dfrac{1}{\sqrt{\bigintss_{-a}^0 \left(- e_{x,n}^2 + e_{z,n}^2\right) dx}} \\
     &= \dfrac{k \epsilon_{f,n}}{\sqrt{-\dfrac{k^2 \epsilon_{f,n} a}{2} + \left(\dfrac{\alpha_{f,n}^2 - \beta^2}{2 \alpha_{f,n}} \right)\sin\left(\alpha_{f,n}a\right) \cos\left(\alpha_{f,n} a - 2 \phi_n^\textrm{TM}\right)}}.
\end{split}
\end{equation}

\subsubsection{$\beta$ modes}
Similarly, for TE modes with $\beta$ eigenvalue, the electric field modal profile is given by 
\begin{equation}
    \bv{E}_n(x,z) = \hat{y} \, A_n^{\textrm{TE}} \,u_n(x)  \, e^{i \beta_n z},
    \label{eq:E_mode}s
\end{equation}
where
\begin{equation}
u_n(x) = \begin{cases}
    \cos(\phi_n^{\textrm{TE}})\, e^{i\alpha_{c,n} x}, & \text{for}\ x > 0, \\
    \cos\left(\alpha_{f,n} x + \phi_n^{\textrm{TE}} \right), & \text{for}\ -a \leq x \leq 0, \\
    \cos\left(\alpha_{f,n} a - \phi_n^{\textrm{TE}} \right) \; e^{-i\alpha_{s,n} (x + a)}, & \text{for}\ x < -a,
    \end{cases}
\end{equation}
and
\begin{equation}
\phi_n^{\textrm{TE}} = \tan^{-1}\left( \dfrac{-i\alpha_{c,n}}{\alpha_{f,n}}\right).
\end{equation}}
{\XYZ The corresponding magnetic field vector for TE modes will be
\begin{equation}
\bf{H}_n(x,z) = A_n^{\textrm{TE}} \bf{h}_n(x) e^{i \beta_n z},
\end{equation}
where the components of the transverse spatial variation vector, $\bf{h}_n(x)$, in regions $j \in \{c,f,s\}$ is related to $u_n(x)$ as follows:
\begin{equation}
h_{x,n} =  -\dfrac{\beta_n}{k \mu_j} u_n(x) ,\quad h_{z,n} = -\dfrac{i}{k \mu_j} \dfrac{\partial u_n(x)}{\partial x}.
\end{equation}
Here, the index $n$ represents the $n^{th}$ eigenvalue of $\beta$, and the amplitude $A_n^{\textrm{TE}}$ is the normalization constant of the $n^{th}$ mode, given by
\begin{equation}
\begin{split}
    A_n^{\textrm{TE}} &= \dfrac{1}{\sqrt{-i \left[h_{z,n} \dfrac{\partial u_n}{\partial \beta_n} - u_n \dfrac{\partial h_{z,n}}{\partial \beta_n} \right]_{x = -a^-}^{x=0^+} + 2 \bigintss_{-a}^0 u_n(x) h_{x,n} \, dx }}, \\
    &= \dfrac{1}{\sqrt{ - \dfrac{i \beta_n}{k \mu_c \alpha_{c,n}} u_n^2(0)  - \dfrac{i \beta_n}{k \mu_s \alpha_{s,n}} u_n^2(-a) - \dfrac{2 \beta_n}{k \mu_f}\bigintss_{-a}^0 u_n^2(x) \, dx }}.
    \end{split}
\end{equation}

For TM modes, the mode profile is defined in terms of the $y$-polarized magnetic field, i.e.,
\begin{equation}
    \bv{H}_n(x,z) = \hat{y} \, A_n^{\textrm{TM}} \, v_n(x)\, e^{i \beta_n z},
     \label{eq:H_mode}
\end{equation}
where
\begin{equation}
v_n(x) = \begin{cases}
     \cos(\phi_n^{\textrm{TM}})\, e^{i\alpha_{c,n} x}  , & \text{for}\ x > 0, \\
    \cos\left(\alpha_{f,n} x + \phi_n^{\textrm{TM}} \right), &  \text{for}\ -a \leq x \leq 0, \\
    \cos\left(\alpha_{f,n} a - \phi_n^{\textrm{TM}} \right) \; e^{-i\alpha_{s,n} (x + a)} , & \text{for}\ x < -a,  \\
    \end{cases}
\end{equation}
and
\begin{equation}
\phi_n^{\textrm{TM}} = \tan^{-1}\left[ \dfrac{-i\epsilon_f \, \alpha_{c,n}}{\epsilon_c \, \alpha_{f,n}}\right].
\end{equation}
The corresponding electric field vector for TM modes will be
\begin{equation}
\bf{E}_n(x,z) = A_n^{\textrm{TM}} \bf{e}_n(x) e^{i \beta_n z},
\end{equation}
where the components of the transverse spatial variation vector, $\bf{e}_n(x)$, in regions $j \in \{c,f,s\}$ are related to $v_n(x)$ as follows:
\begin{equation}
e_{x,n} =  \dfrac{\beta_n}{k \epsilon_j} v_n(x) ,\quad e_{z,n} = \dfrac{i}{k \epsilon_j} \dfrac{\partial v_n(x)}{\partial x}.
\end{equation}
The normalization factor $A_n^{\textrm{TM}}$ is given by
\begin{equation}
\begin{split}
 A_n^{\textrm{TM}} &= \dfrac{1}{\sqrt{-i \left[-v_{n} \dfrac{\partial e_{z,n}}{\partial \beta_n} + e_{z,n} \dfrac{\partial v_{n}}{\partial \beta_n} \right]_{x= -a^-}^{x=0^+} + 2 \bigintss_{-a}^0 e_{x,n}(x) v_{n}  \, dx }},\\
 &= \dfrac{1}{ \sqrt{ \dfrac{i \beta_n}{k \epsilon_c \alpha_{c,n}} v_n^2(0) + \dfrac{i\beta_n}{k \epsilon_s \alpha_{s,n}} v_n^2(-a)  + \dfrac{2 \beta_n}{k \epsilon_f}\bigintss_{-a}^0   v_n^2(x)  \, dx }}.
 %
\end{split}
\end{equation}}

\section{Permittivity Normal modes of single sphere}
{\XYZ In this Appendix, we present the analytical derivatives of the transcendental equations for a single spherical particle, which are used in our root-search algorithm, along with the corresponding mode field profiles and associated normalization constants for both TM and TE polarizations of the eigenpermittivity roots.

\subsection{Derivatives of transcendental equation}

In this subsection, we define the explicit expressions for the
transcendental equations,
\begin{equation}
\dfrac{\partial f_l^\textrm{(\textrm{TM})}(x)}{\partial x} = -\frac{j_\ell(x) j_\ell'(x) + x j_\ell(x)^2 \left( 1 - \frac{\ell(\ell+1)}{x^2} \right) + x \left( j_\ell'(x) \right)^2}{j_\ell(x)^2},
\end{equation}
and
\begin{equation}
\dfrac{\partial f_l^\textrm{(TE)}(x)}{\partial x} = \frac{-2 j_\ell(x) j_\ell'(x) - j_\ell'(x) \left( j_\ell(x) + x j_\ell'(x) \right)}{x^2 j_\ell(x)^2} + \frac{\ell(\ell+1)}{x^3}  - \dfrac{1}{x} - \frac{2}{x^3}.
\end{equation}
In this case, the analytical derivative of the transcendental equations with respect to $x$ is simple (no need to apply the chain rule).

\subsection{Mode field profiles}

We now define the explicit expressions for the modes of a sphere using the radial function and the two basic types of the vector spherical harmonics (VSHs), transverse electric (TE) and transverse magnetic (TM) modes, i.e.,
\begin{equation}
\begin{split}
TE: \vec{E}^{(\textrm{TE})}_{nlm}(\vec{r}) =& f^{(\textrm{TE})}_{nl}(r) \vec{X}_{lm}(\theta,\phi), \\
TM: \vec{E}^{(\textrm{TM})}_{nlm}(\vec{r}) =& \frac{i}{k\left[1 - u_{nl}^{(\textrm{TM})}\theta(r)\right]} \left[\nabla \times \left(f^{(\textrm{TM})}_{nl}(r) \vec{X}_{lm}(\theta,\phi)\right)\right],
\end{split}
\label{eq:M_E_multipoles}
\end{equation}
where $\vec{X}_{lm}(\theta,\phi)$ is the vector spherical harmonics, and the radial function $f^{(\textrm{F})}_{nl}$ for both $F = \textrm{TE},\textrm{TM}$, takes the form
\begin{equation}\label{eq:flnF}
f_{nl}^{(F)}(r) = \left\{
\begin{array}{cc}
A_{nl}^{(F)} j_l\left(k \left[1 - u^{(F)}_{nl}\right]^{1/2} r\right), & r < a, \\
B_{nl}^{(F)} h^{(1)}_l(k_b r), & r > a.
\end{array} \right.,
\end{equation}
where $u_{nl}^{(F)} \equiv 1 - \epsilon_{nl}$. The continuity of $f^{(\textrm{TE})}_{nl}$ ensures continuity of $H_\perp = H_r$ and the continuity of $\frac{d f_{nl}^{(\textrm{TE})}}{d r}$ ensures the continuity of $H_\parallel$.


Continuity of $f_{nl}^{(F)}$ dictates that
\begin{equation}
B_{nl}^{(F)} = A_{nl}^{(F)} \frac{j_l^{(1)}\left(x_{nl}^{(F)}\right)}{h_l^{(1)}(x_b)},
\end{equation}
where
\begin{equation}
x_{nl}^{(F)} \equiv k \left(1 - u_{nl}^{(F)}\right)^{1/2} a. \label{eq:xlnF}
\end{equation}
Now, demanding continuity of the parallel electric and magnetic field components, we get Eq.~(\ref{eq:TM_roots_search}) and Eq.~(\ref{eq:TE_roots_search}) from which we can determine the eigenvalues.

It is important to note that the dispersion relations depend on $k a = a \omega / c$, so that computing the eigenvalues for different particle sizes is equivalent to computing them for different frequencies.

\subsection{Modes normalization}
The coefficients $A_{nl}^{(F)}$ are determined from the normalization condition on $\bf{E}_{nlm}^{(F)}$. For $F = TE$,
\begin{equation}
\begin{split}
1 &= \langle C \bf{E}_{nlm}^{(\textrm{TM})} | \bf{E}_{nlm}^{(\textrm{TM})} \rangle= \int_{r<a} d^3 r \left[A_{nl}^{(\textrm{TM})} j_l\left(k \left(1 - u_{l,n}^{(\textrm{TM})}\right)^{1/2}r\right)\right]^2 \bf{H}_{lm}^* \cdot \bf{H}_{lm},\\
&= \left(A_{nl}^{(\textrm{TM})}\right)^2 \int_0^a r^2 dr \left[j_l\left(k \left(1 - u_{l,n}^{(\textrm{TM})}\right)^{1/2}r\right)\right]^2= \frac{\left(A_{nl}^{(\textrm{TM})}\right)^2 a^3}{\left(x_{nl}^{(\textrm{TM})}\right)^3} \int_0^{x_{nl}^{(\textrm{TM})}} dx x^2 \left[j_l(x)\right]^2,\\
&=\frac{\left(A_{nl}^{(\textrm{TM})}\right)^2 a^3}{2\left(x_{nl}^{(\textrm{TM})}\right)^2} \bigg(\left[\psi'_l\left({x_{nl}^{(\textrm{TM})}}\right)\right]^2 +  \left[\psi_l\left({x_{nl}^{(\textrm{TM})}}\right)\right]^2-  l(l + 1) j_l^2\left({x_{nl}^{(\textrm{TM})}}\right) - \\
& \qquad \qquad \qquad \qquad j_l\left(x_{nl}^{(\textrm{TM})}\right)\psi'_l\left({x_{nl}^{(\textrm{TM})}}\right)\bigg).
\end{split}
\label{eq:norm_A_M_gen}
\end{equation}
Here, $\psi_l(x)\equiv xj_l(x)$. Since $\bf{H}_{lm}$ is a Hermitian operator, its eigenstates, namely the VSHs, satisfy the usual orthogonality relations when integrated over the angular variables ~\cite{Bergman_Stroud_PRB_1980}.\\
while for $F = TM$,
\begin{eqnarray}
1 &=& \langle C \bf{E}_{nlm}^{(\textrm{TE})} | \bf{E}_{nlm}^{(\textrm{TE})} \rangle \nn \\
&=& \left(\frac{i A_{nl}^{(\textrm{TE})}}{k \left(\epsilon_{nl}^{(\textrm{TE})}\right)}\right)^2 \int_{r<a} d^3 r \left[\nabla \times j_l\left(k\left(\epsilon_{nl}^{(\textrm{TE})}\right)^{1/2} r\right) \bf{H}_{lm}^*\right] \cdot \left[\nabla \times j_l\left(k\left(\epsilon_{nl}^{(\textrm{TE})}\right)^{1/2}r\right) \bf{H}_{lm}\right] \nonumber \\
&=& \frac{a^7k^2}{\left({x_{nl}^{(\textrm{TE})}}\right)^7 } \left(A_{nl}^{(\textrm{TE})}\right)^2 \int_0^{x_{nl}^{(\textrm{TE})}} \left(x^2 \left[l(l+1) + 1\right] j^2_l(x) + \left[x j'_l(x) j_l(x) \left(2 + \frac{x j'_l(x)}{j_l(x)}\right)\right]\right) dx. \nonumber
\end{eqnarray}
Thus,
\begin{equation}
\begin{split}
1 = \frac{a^5k^2\left(A_{nl}^{(\textrm{TE})}\right)^2}{2\left(x_{nl}^{(\textrm{TE})}\right)^4} & \bigg(
\left[\psi'_l \left({x_{nl}^{(\textrm{TE})}}\right)\right]^2 +  \left[\psi_l\left({x_{nl}^{(\textrm{TE})}}\right)\right]^2  \\ &- l(l + 1)j_l^2 \left({x_{nl}^{(\textrm{TE})}}\right)  + j_l\left(x_{nl}^{(\textrm{TE})}\right)\psi'_l\left({x_{nl}^{(\textrm{TE})}}\right)\bigg).
\end{split}
\label{eq:norm_A_E_gen}
\end{equation}

We note that the orthogonalization integrals extend to the sphere volume only. The Bessel function integrals coincide with the available formula for integrals over the Bessel functions below~\cite{mcphedran2018killing}. We get 
\begin{eqnarray}
&&\int_0^{x_{nl}^{(F)}} y^2 j_l^2(y) dy = \\ && = \frac{{x_{nl}^{(F)}}\left(\left[\psi'_l\left({x_{nl}^{(F)}}\right)\right]^2 + \left[\psi_l\left({x_{nl}^{(F)}}\right)\right]^2 -l(l + 1)j_l^2\left({x_{nl}^{(F)}}\right)-j_l\left(x_{nl}^{(F)}\right)\psi'_l\left({x_{nl}^{(F)}}\right)\right)}{2}, \nonumber\\
&&\int_0^{x_{nl}^{(F)}} \left[\psi'_l(y)^2 + l(l+1)j_l^2(y)\right] dy = \\ &&=\frac{{x_{nl}^{(F)}}\left(\left[\psi'_l\left({x_{nl}^{(F)}}\right)\right]^2 + \left[\psi_l\left({x_{nl}^{(F)}}\right)\right]^2-l(l + 1) j_l^2\left({x_{nl}^{(F)}}\right) + j_l\left(x_{nl}^{(F)}\right)\psi'_l\left({x_{nl}^{(F)}}\right)\right)}{2}. \nonumber
\end{eqnarray}
An alternative expression for these integrals is given in~\cite{Forestiere2016}, which is expected to be numerically equivalent.}





\bibliographystyle{elsarticle-num}







\end{document}